\documentstyle[12pt,fleqn,cite,epsfig]{article}

\begin{document}
\topmargin 0pt
\oddsidemargin 0mm

\begin{titlepage}
\begin{flushright}
IP/BBSR/2002-08\\
hep-th/0204025
\end{flushright}

\vspace{5mm}
\begin{center}
{\Large \bf D-Brane Solutions in pp-wave Background}
\vspace{6mm}

{\large
Alok Kumar, Rashmi R. Nayak and Sanjay}\\
\vspace{5mm}
{\em Institute of Physics\\ 
Bhubaneswar 751 005\\ India\\
\vspace{3mm}

email: kumar, rashmi, sanjay@iopb.res.in}

\end{center}
\vspace{5mm}
\centerline{{\bf{Abstract}}}
\vspace{5mm}
We present classical solutions for a $D5$ and $NS5$-branes in  
a pp-wave background. The worldvolume coordinates 
for these branes lie along a six dimensional 
pp-wave configuration obtained from $AdS_3 \times S_3$ in a Penrose limit. 
One in addition has nontrivial R-R 3-form field strength as well as
a dilaton background. Classical solutions for $Dp$-branes ($p> 5$), 
as well as their bound states, are then 
generated in both IIA and IIB string theories by using $T$-duality
symmetries. We also briefly discuss the construction of $Dp$-branes in 
pp-wave backgrounds, obtained from string theories on $AdS_m\times S^m$. 
Finally, the construction of the above
D5-brane solution is presented from the point of view of the Green-Schwarz 
worldsheet string theory and its low lying spectrum is discussed.

\end{titlepage}

\newpage

String theory in the presence of pp-wave backgrounds 
\cite{gueven,amati,tseytlin,seng,seng2,seng3,guev2} has generated 
lots of excitement among
physicists searching for a deep connection between quatum gravity and 
gauge theory dynamics. Such string backgrounds  
are technically tractable, and have direct 
applications to the four dimensional conformal theories from the point 
of view of a duality between string and gauge theories.
The fact that 
pp-waves provide exact solutions of string theory, including in the 
presence of $R-R$ fields\cite{seng}, is now known for long time. It has
now been realized\cite{matsaev,russo} 
that in some special cases, they give rise to exactly solvable 
string theory with a quadratic action. As a result the spectrum of 
the theory can be explicitly obtained.  
Moreover, an interesting connection of pp-wave background
with gauge field theories emerges by realizing that they 
can be obtained\cite{blau} in Penrose\cite{penrose} scaling limit 
from a geometry of  $AdS_p \times S^q$ type.  
This finally leads to a duality between string theories in 
pp-wave geometries and gauge theories, originally living 
on the boundary of the $AdS$ spaces\cite{malda,mukhi,sonn,mohsen,martin}. 

In view of the importance of BPS branes (preserving 
certain amount of supersymmetries)
in understanding the 
non-perturbative dynamics of string theories, 
in this paper, we first present classical solutions of a $D5$
and an $NS5$-brane in the
background of a pp-wave. These are obtained from known 
classical solutions for such branes
in $AdS_3 \times S^3$ background\cite{papa}.
By setting the brane charges to zero, 
in our case, we obtain a ten dimensional solution with a six dimensional 
pp-wave background, giving rise to a massive Green-Schwarz string 
theory. Four spatial dimensions transverse to the branes, for 
these solutions, are flat. 

To obtain our solution, 
we now begin by writing the $1_{NS}$ + $5_{NS}$ + $5'_{NS}$ solution
given in \cite{papa}:
\begin{equation}
ds^2 = g_1^{-1}(x,y)(-dt^2 + dz^2) + H_5(x) dx^n dx^n 
+ H'_5(y) dy^m dy^m, \label{g155}
\end{equation}
\begin{equation}
dB = dg_1^{-1}\wedge dt \wedge dz + *dH_5 + *dH'_5,
\label{b155}
\end{equation}
\begin{equation}
e^{2\phi} = {H_5(x)H'_5(y)\over g_1(x,y)}, 
\label{d155}
\end{equation}
where
\begin{equation}
[H'_5(y) {\partial_x}^2 + H_5(x) {\partial_y}^2] g_1(x,y)=0.
\end{equation}
A paticular solution for $g_1$ is given as: 
\begin{equation}
g_1(x,y) = H_1(x)H'_1(y), 
\end{equation}
with
\begin{equation}
H_1 = 1 + {R_1^2\over x^2},~~~H_5 = 1 + {R_5^2\over x^2},~~~
H'_1 = 1 + {{R'}_1^2\over y^2},~~~H'_5 = 1 + {{R'}_5^2\over y^2}.
\end{equation}
By setting $H'_1 = 1$, the above solution becomes a direct product
of NS1 + NS5 configuration, together with an NS5-brane. Then, in a 
decoupling limit $x \rightarrow 0$, one obtains from the metric 
given in (\ref{g155}) \cite{papa}: 
\begin{equation}
ds^2 = {x^2\over R_1^{2}}(-dt^2 + dz^2) + R_1^{2} {dx^2\over x^2} +
R_1^{2}d\Omega_3^{2} + H'_5(y)(dy^2 + y^2 d{\Omega'_3}^{2}),
\label{ns33}
\end{equation}
where one has also used $H_1 = H_5$ to simplify the expression. 
Finally, by using an S-duality transformation on the background
mentioned above, one obtains a metric:
\begin{equation}
ds^2 = {H'_5}^{-1/2} \Big[{x^2\over R_1^{2}}(-dt^2 + dz^2) + R_1^{2} 
{dx^2\over x^2} + R_1^{2}d\Omega_3^{2} \Big] + {H'_5}^{1/2}(y)(dy^2 + y^2
d{\Omega'_3}^{2}).
\label{d33}
\end{equation}

We now obtain the $D5$ brane solution in the pp-wave background, 
by interpreting the metric in eqn.(\ref{d33})
as that of a $D5$-brane in $AdS_3 \times S_3 \times R^4$.
The coordinates $t, x$ and $z $ in the above solution 
represent the $AdS_3$ space, whereas $d \Omega_3^2 $ is the
metric of an $S^3$. It is also evident that the worldvolume of the 
above $D5$-brane solution coincides with this $AdS_3 \times S^3$
space and transverse directions to the branes are represented by 
the coordinates $y^m$. 
We now apply the Penrose limit\cite{blau} on $t, x, z$, as well 
as on the coordinates of $S^3$ parametrized as: 
\begin{equation}
 d\Omega_3^2 = d\psi^2 + sin^2 \psi d\Omega_2^2,
\label{gs3}
\end{equation}
on the $D5$-brane metric in (\ref{d33}). Then, by following the 
procedure in \cite{blau}, one obtains from the metric in 
(\ref{d33}):
\begin{equation}
dS^2 = {H'_5}^{-1/2} [ du dv - \mu^2\sum_{i=1}^4 {z^i}^2 du^2 + 
\sum_{i=1}^4 dz^i dz^i ] + {H'_5}^{1/2} 
(dy^2 + y^2 d{\Omega'_3}^2).\label{dpp}  
\end{equation}
As is noticed, the $y$ dependence in the limiting procedure
described above remains unchanged due to a simultaneous scaling of 
$y$ together with $R'_5$.

The expression for the dilaton field in the Penrose limit above
can be found rather easily and is given by 
\begin{equation}
e^{2 \phi} = {{H'_5}(y)}^{-1}. \label{d5dilaton}
\end{equation}

The expression for the ($R-R$) (3rd rank antisymmetric) field strength ($H^R$)
in the pp-wave limit is obtained from the NS-NS field strength 
given in expression (\ref{b155}) by applying S-duality
transformation on these fields. 
As is noticed, in this case there are 
several nonzero components. In the pp-wave limit, the constant field
strengths: $H^R_{+12}$ and $H^R_{+34}$ emerge from the first two terms
in the RHS of eqn. (\ref{b155}), in two steps. First one obtains, 
in $x\rightarrow 0$ decoupling limit, expression 
for the field strength that is given by the  
appropriate levi-civita tensors on $AdS_3$\cite{ali} and $S^3$ respectively. 
In the second step, one applies the Penrose limit following the procedure 
mentioned above to get constant field strengths\cite{malda}: 
\begin{equation}
H^R_{+ 12 } =  H^R_{+34} = 2 \mu \label{33}. \label{h1234} 
\end{equation}
We however notice that the last term in the RHS of eqn.
(\ref{b155}) also survives the above limiting processes and its functional
form remains unchanged. As a result we have additional nonzero components of 
the $R-R$ 3-form field strength\cite{tset2}:
\begin{equation}
H^R_{m n p} = \epsilon_{m n p l} \partial_l H'_5 (y). \label{hijk}
\end{equation} 
Here we once again like to emphasize that all the three terms:
representing the field strengths on $AdS_3$, $S^3$ and $R^4$,
of a $D5$ brane solution, 
have identical scaling properties, when standard Penrose scalings of
\cite{blau} are applied in the $AdS_3 \times S^3$ part and transverse
coordinates are scaled appropriately. 
We have therefore presented a classical solution corresponding to 
a $D5$-brane in an exactly solvable pp-wave background. 

We can also write down the NS5 brane solution in a pp-wave background
using the metric in equation (\ref{ns33}). One now obtains the metric:
\begin{equation}
dS^2 = [ du dv - \mu^2\sum_{i=1}^4 {z^i}^2 du^2 + 
\sum_{i=1}^4 dz^i dz^i ] + {H'_5} (dy^2 + y^2 d{\Omega_3}^2).  
\end{equation}
The dilaton now has a form:
\begin{equation}
e^{2 \phi} = {{H'_5}(y)}.
\end{equation}
The antisymmetric tensors are exactly of the form in 
eqns. (\ref{h1234}) and (\ref{hijk}), 
with `R' replaced by `NS' in these equations. 

Now we generate $Dp$-branes ($p>5$),
starting with our solution as
given in  eqns.(\ref{dpp})- (\ref{hijk}) by applying 
T-duality\cite{ortin,myers} 
along the transverse directions of the $D5$-brane. Applying this
procedure to a tranverse coordinate $\tilde{x}$ one obtains a 
$D6$ brane in type IIA theory. The explicit solution  
for the non-zero components of the metric, 
field strengths and dilaton are then given as follows: 
\begin{eqnarray}
e^{2\phi_a} &=& H_6^{-3/2},\cr
& \cr
A^{(1)}_{m} &=& A^{(R)}_{\tilde x m}: \>\>\>                     
F^{(1)}_{m n} = \epsilon_{m n p}\partial_{p}H_6,\cr
& \cr
A^{(3)}_{\tilde x\mu\nu} &=& A^{(R)}_{\mu\nu},\>\>\>\> 
\tilde{F}_{\tilde x\alpha\beta\gamma} = {\partial_\gamma}A_{\tilde 
x\alpha\beta} + cycl.  
= H^{(R)}_{\alpha\beta\gamma}:  \>\>\>
\tilde{F}_{\tilde x u 1 2 } = \tilde{F}_{\tilde x u 3 4 } = 2 \mu,
\label{fd6pp}
\end{eqnarray}
\begin{equation}
dS^2 = {H_6}^{-1/2} [ du dv - \mu^2\sum_{i=1}^4 {z^i}^2 du^2 + 
\sum_{i=1}^4 dz^i dz^i + d\tilde x^2 ] + 
{H_6}^{1/2} (dr^2 + r^2 d{\Omega_2}^2),
\label{d6pp}  
\end{equation}
with $H_6 = ( 1 + R_6/r)$, $r$ being the radial coordinate in 
transverse direction and $H^{(R)}$ denotes the field strength
that we found earlier for the $D5$ brane solution in the IIB theory.
Integer subscripts in field strenghts above denote coordinates 
$z^i$'s and indices $\mu, \nu$ run over the longitudinal directions.
For $\mu=0$ this solution reduces to the standard $D6$ brane in 
type IIA theories\cite{phys.rep.}. 
One can continue this exercise further to the higher branes 
as well. Then, for example, a logarithimic Green function appears
for a $D7$ brane and a linear function appears in 
the solution for a $D8$ brane\cite{berg}. 
The explicit solution for the $D7$ brane 
($ H_7 = 1 - \mu \ln (r/l))$\cite{myers} is given as:
\begin{eqnarray}
e^{2\phi_b} &=& {H}^{-2}_{7},\cr
& \cr 
\chi & =& - A^{(1)}_{\tilde y}: \>\>\>
\partial_{m} \chi = - \epsilon_{m n}\partial_n H_7,\cr 
& \cr
F^{(5)}_{\tilde{x} \tilde{y} u 1 2} & = & 
F^{(5)}_{\tilde{x} \tilde{y} u 3 4} = 2 \mu,  
\end{eqnarray}
\begin{equation}
dS^2 = {H_7}^{-1/2} [ du dv - \mu^2\sum_{i=1}^4 {z^i}^2 du^2 + 
\sum_{i=1}^4 dz^i dz^i + d\tilde x^2 + d\tilde{y}^2
 ] + {H_7}^{1/2} (dr^2 + r^2 d{\theta}^2).\label{d7pp}  
\end{equation}
Solutions corresponding to the bound states of $Dp$ branes can also be 
constructed using the procedure given in \cite{myers}. For example, a
$D5-D7$ bound state solution can be obtained from the $D6$ brane
solution in eqns.(\ref{fd6pp}) and (\ref{d6pp}) by 
applying a rotation between the longitudinal coordinate $\tilde{x}$ and 
one of 
the transverse coordinate $\tilde{y}$ for this solution, and then 
applying T-duality along (rotated)
coordinate $\tilde{y'}$. We skip the details
of this solution and now briefly discuss the construction of $Dp$
branes in other pp-wave backgrounds. These backgrounds can be  
obtained from string theory on $AdS_m \times S^m$ for some other
values of $m$.   In $m=2$ case, for example, relevant field strength
living on $AdS_2$ and $S^2$ now correspond to a $1$-form field. It should 
therefore be possible to constuct $D4$ branes in a 
pp-wave background, 
now originating from $AdS_2\times S^2\times R$. One now has $R^5$ as a
five dimensional transverse space on which a $4$-form field strength
of the type: $F_{m n p q} = \epsilon_{m n p q r} \partial_r H_4$
lives. One in addition will have constant field strengths of the type:
$F^{(1)}_{+ 1} = F^{(1)}_{+ 2} = \mu $.
Once again, by applying $T$-dualities, all the higher branes can 
be obtained. This process can also be repeated for $D8$ branes in 
a pp-wave bakground obtained from $AdS_4\times S^4\times R$. We hope
to come back with explicit solutions of these types in near future.

We now proceed to construct the $D5$-brane, obtained earlier as a
classical solution in supergravity, from the point of view
of a first quantized string theory in Green-Schwarz formalism in light-cone
gauge. The relevant classical action to study in our case, obtained 
from $AdS_3 \times S^3 \times R^4$ by applying the Penrose limit, 
has a form\cite{russo}:
\begin{eqnarray}
L &=& L_B + L_F,\label{t-action}\cr
& \cr
L_B &=& \partial_{+}u\partial_{-}v - m^{2} x_i^{2} +
\partial_{+}x_i\partial_{-}x_i +
\partial_{+}x_{\alpha}\partial_{-}x_{\alpha}.
\label{baction}
\end{eqnarray}
\begin{equation}
L_F = i\theta_R\gamma^{v}\partial_{+}\theta_R +
i\theta_L\gamma^{v}\partial_{-}\theta_L -
2im\theta_L\gamma^{v}M\theta_R,
\label{faction}
\end{equation}
\begin{equation}
m \equiv \alpha'p^{u} \mu = 2 \alpha'p_{v} \mu,
\end{equation}
\begin{equation}
M = -{1\over 2}(\gamma
^{1 2} + \gamma ^{3 4}),
\label{projM}
\end{equation}
where $(\theta_R, \theta_L)$ are 16-component Majorana-Weyl spinors in 
the right and left moving sectors. $x_i$ and $x_{\alpha}$ in 
eqn.(\ref{baction}) refer to the longitudinal $D5$ brane coordinates
$z^i$ and transverse ones $y^m$ in the metric given in eqn.(\ref{g155}).
The form of the matrix $M$ in the 
present case leads to only half of the fermions picking up 
masses\cite{russo}. This is different from the case of 
pp-wave background that is obtained from $AdS_5 \times S^5$
\cite{dabh,lee},
and leads to some differences in our case compared to the 
one in \cite{dabh}.

The equations of motion following from (\ref{t-action}) has a form:
\begin{equation}
\partial_{+}\partial_{-} x_i + m^2 x_i =
0,~~~~~~~~\partial_{+}\partial_{-} x_{\alpha} = 0, \label{beqn}
\end{equation}
\begin{equation}
\partial_{+}\theta_R - m M \theta_L = 0,~~~~~~~~ \partial_{-}\theta_L
- m M \theta_R = 0.  \label{feqn}
\end{equation}
After solving the light-cone gauge condition one rewrites the above
action, as well as equations of motion using 8-component 
spinors $S_L$ and $S_R$. The equations of motion have an identical 
form as in (\ref{beqn}), (\ref{feqn}) by replacing $\theta_{L,R}$
by $S_{L,R}$. The matrix $M$ now is an $8\times 8$ one of the same 
structure as in (\ref{projM}).
Keeping in mind the explicit form of the matrix $M$, which evidently 
breaks the $SO(8)$ further, 
one splits the fermions in the $8\rightarrow 4 + 4$ way: 
$S_L\rightarrow (\tilde{S}_L,{\hat{S}_L})$, 
$S_R\rightarrow (\tilde{S}_R,{\hat{S}_R})$:
\begin{eqnarray}
\gamma^{1234} \pmatrix{\tilde{S}_{L,R} \cr \hat{S}_{L,R}} = 
\pmatrix{- \tilde{S}_{L,R} \cr \hat{S}_{L,R}}. \label{ga1234}
\end{eqnarray}
Finally one introduces $4\times 4$ matrices $\Lambda$ and $\Sigma$:
\begin{eqnarray}
\gamma^{12}\pmatrix{\tilde{S}_{L,R} \cr \hat{S}_{L,R}} = 
-\pmatrix{\Lambda \tilde{S}_{L,R} \cr\Sigma \hat{S}_{L,R}}, \label{gamma12}
\end{eqnarray}
with $\Lambda^2 = \Sigma ^2 = -1$. 
$\Lambda$ and $\Sigma$ in the above equation are $4\times 4$ 
antisymmetric matrices with eigenvalue $\pm i$. Using these 
notations one has:
\begin{eqnarray}
M \pmatrix{\tilde{S}_{L,R} \cr \hat{S}_{L,R}} 
= \pmatrix{\Lambda \tilde{S}_{L,R} \cr  0}.
\end{eqnarray}
The equations of motion written in terms of $(\tilde{S}_L,{\hat{S}}_L)$ and 
$(\tilde{S}_R,{\hat{S}}_R)$ are then of the form: 
\begin{equation}
\partial_{+}\tilde{S}_{R} - m \Lambda \tilde{S}_{L}=0, \>\>\>\>
\partial_{-}\tilde{S}_{L} - m \Lambda \tilde{S}_{R}=0,
\end{equation}
\begin{equation}
\partial_{+}\hat{S}_{R} =0, \>\>\>\>
\partial_{-}\hat{S}_{L} =0.
\end{equation}

We now write down the open string
boundary condition\cite{dabh} corresponding to the 
above $D5$-brane configuration.
These are given on the longitudinal bosonic coordinates $x^{i}$, 
($i=1,..,4$) and transverse ones $x^{\alpha}$ ($\alpha = 5,..,8$) by 
standard Neumann and Dirichlet conditions. For spinors we have:
\begin{equation}
{S_L}|_{\sigma = 0, \pi} = \Omega S_R |_{\sigma = 0, \pi}, \label{boundary}
\end{equation}
with $\Omega$ being the product of $\gamma$ matrices along the 
Dirichlet directions. $\Omega$ has to satisfy the following 
consistency conditions in the present case:
\begin{equation}
 [\Omega , \gamma] = 0, 
\end{equation}
\begin{equation}
 \Omega M \Omega = M. 
\end{equation}
The first of these equations follows from the requirement that $S_{L}$ 
and $S_R$ have the 
same $SO(8)$ chirality. The second condition is a consequece 
of a consistency condition on the above equations of motion in the 
zero mode sector, leading to one half supersymmetry. 
The explicit form of $\Omega$ in our case is given 
by $\Omega = \gamma^{5678}$ and leads to the following boundary conditions on 
$(\tilde{S}_L,{\hat{S}}_L)$ and $(\tilde{S}_R,{\hat{S}}_R)$ 
using eqn.(\ref{ga1234}):
\begin{equation}
\tilde{S}_L|_{\sigma = 0, \pi} = - \tilde{S}_R |_{\sigma = 0, \pi}, 
\label{boundary2}
\end{equation}
\begin{equation}
\hat{S}_L |_{\sigma = 0, \pi} = \hat{S}_R |_{\sigma = 0, \pi}. 
\label{boundary2'}
\end{equation}
Now, as the equations of motion as well as the boundary condition 
for the components $\hat{S}_{L,R}$ are identical to the
ones in flat space, we only concentrate on finding explicit solution for 
$\tilde{S}_{L,R}$ below. 

Explicit solution for $x^i$ and $x^{\alpha}$ are given as:
\begin{equation}
x^i(\sigma,\tau) =  [{x_0^i}
cos m \tau + {1\over m} p_0^i sin m \tau] + i \sum_{n \neq 0}
{1\over\omega_n} {{\alpha_n^i}} e^{-i \omega_n \tau} cos n\sigma,
\label{xi}
\end{equation}
\begin{equation}
{x^\alpha}(\sigma,\tau) = \sum_{n\neq 0} {1\over n} 
{\alpha_n^\alpha}
e^{-i n\tau} sin {n\sigma}, \label{soln1-b}
\end{equation}
where $\omega_n = sgn(n)\sqrt{n^2 + m^2}$ and 
oscillators satisfy the following commutation relations: 
\begin{equation}
[a^I_n,a^J_l] =\delta_{n+l} \delta^{IJ} ,\>\>\> 
[\bar{a}^i_0,a^j_0] = \delta^{i j}, 
\end{equation}
with indices $I, J$ running over all the eight coordinates. We
have also used the definitions:
\begin{equation}
a_0^i = {1\over \sqrt{2 m}} (p_0^i + i m x_0^i),\>\>\> 
\bar{a}_0^i = {1\over \sqrt{2 m}} (p_0^i - i m x_0^i),\>\>\>
a^I_n = \sqrt{1\over{|\omega_n|}} \alpha^I_n.
\end{equation}

For $\tilde{S}_{L,R}$ we have the solution of the equations of motion
\cite{russo}:
\begin{equation}
\tilde{S^a}_L =    [\tilde {S^a}_0   cos\omega_0\tau
-\bar{\tilde {S^a}_0} sin \omega_0\tau]
+\sum_{n\neq 0} {c_n\over \sqrt 2} 
[\tilde\phi{_n}\tilde {S^a}{_n} + {i\over m}
(\omega_n - n)\Lambda^a_b {S^b}_n \phi_n ], 
\label{solsl}
\end{equation}
\begin{equation}
\tilde{S^a}_R = \Lambda^a_b 
[\bar{\tilde{S}}^b_0 cos \omega_0\tau
+ {\tilde {S^b}_0} sin \omega_0\tau ]
+\sum_{n\neq 0} {c_n \over \sqrt 2}
[\phi{_n} {S^a}{_n} +  {i\over m}(\omega_n - n)
\Lambda^a_b {\tilde {S^b}_n}\tilde{\phi_n}], 
\label{solnsr}
\end{equation}
where $c_n = m/\sqrt{m^2 + (\omega_n - n)^2)}$ and 
\begin{equation}
\tilde\phi_n = \exp {-i(\omega_n \tau + n\sigma)}, \>\>\>
\phi_n = \exp {-i(\omega_n \tau - n\sigma)}, \>\>\>
(S_{-n} = S_n^{\dagger} \>\>etc.)\label{phin}
\end{equation}
Moreover, the boundary condition eqn. (\ref{boundary}) implies:
\begin{equation}
\tilde{S^a_0} = - (\Lambda \bar{\tilde S_0})^a, \>\>\>
\tilde{S^a_n} = - S^a_n. \label{sosn}
\end{equation}
One can now put the conditions (\ref{sosn}) back into the solutions 
in eqns. (\ref{solsl}) and (\ref{solnsr}) to find out that indeed the boundary 
conditions in eqns. (\ref{boundary2}) and (\ref{boundary2'})
are being satisfied. 
The commutation relations for the above spinor components turn out to be 
of the form:
\begin{equation}
\{\tilde{S}_0^a, \tilde{S}_0^b \} = \delta^{ab}, \>\>\> 
\{S_n^a,{S_{-n}^b} \} = \delta^{ab}.
\end{equation}

The total Hamitlonian can be written as:
\begin{equation}
H = H_0 + \tilde{H} + \hat{H},
\label{thamilton}
\end{equation}
where
\begin{eqnarray}
H_0 &=& {m\over p^u} (\bar{a}_0^i a_0^i - i \tilde{S^a_0} 
\Lambda \tilde{S^a_0}) + e_0, \cr
& \cr
\tilde{H} &=& {1\over p^u}\sum_{n=1}^{\infty}
\omega_n ({a_n^i}^\dagger a_n^i + 
{S^a_n}^\dagger S^a_n), \cr
& \cr
\hat{H} &=& {1\over p^u}
\sum_{n=1}^{\infty} n ({a_n^{\alpha}}^\dagger a_n^{\alpha} + 
{\hat{S^A_n}}^\dagger \hat{S^A_n}). \label{h0n}
\end{eqnarray}
with $e_0$ being the zero point energy.
The form of the zero mode part of the hamiltonian implies
that the ground state degeneracy of a $D5$-brane spectrum is broken.
To analyze the spectrum of the states, we use the fermionic 
creation and annihilation operators constructed out of spinors:
$\tilde{S}$ and $\hat{S}$. Since both of these
are 4 component objects, each of them give rise to two such creation 
operators and two annihilation opertaors. Then sixteen states
(at a level defined by the vacuum energy $e_0 = 2\mu$\cite{dabh}),
in this representation, are obtained using these operators on 
appropriate Fock vacuum for $\mu = 0$.
On the other hand for $\mu \neq 0$, there is a splitting 
in this spectrum which can now be analyzed by simply taking into account
the $H_0$ part in eqns.(\ref{thamilton}) and (\ref{h0n}) .

We have therefore obtained explicit construction of the open string
describing the $D5$-branes in the pp-wave background presented earlier 
in this paper. It will be interesting to generalize these
results to other D-branes and their bound states, discussed in the paper, 
as well. We however leave this as a future exercise.



\end{document}